\documentstyle{pss}
\input epsf.tex
\newcommand{\Sp}{\mathop{\rm Sp}\nolimits}
\newcommand{\Img}{\mathop{\rm Im}\nolimits}
\begin{document}

\title{Correlated hopping in infinite dimensions:\\
Rigorous local approach}

\author{{\sc A.M. Shvaika}\footnote{) %Corresponding author;
Phone: +380 322 761054, Fax: +380 322 761978, e-mail:
ashv@icmp.lviv.ua} )}

\address{Institute for Condensed Matter Physics of the
National Academy of Sciences of Ukraine,  1~Svientsitskii Str., UA-79011
Lviv, Ukraine}

\submitted{\today} \maketitle

\hspace{9mm} Subject classification: 71.10.Fd, 71.15.Mb, 05.30.Fk,
71.27.+a

\begin{abstract}
The general approach for the description of correlated hopping in the
Dynamical Mean-Field Theory which is based on the expansion over electron
hopping around the atomic limit is developed. It is formulated in terms of
the local irreducible parts (cumulants) of Green's functions and allowed
to calculate thermodynamical functions. As a limit case the
Falicov--Kimball model with correlated hopping is considered.

\end{abstract}

The main idea of the Dynamical Mean-Field Theory (DMFT) is in the local
(single-site) nature of the self-energy
\cite{MetznerVollhardt,DMFTreview}. But this statement violates for the
systems with correlated hopping, when self-energy becomes unlocal
\cite{Schiller}. In this article we present the general DMFT approach for
the description of correlated hopping which is based on the expansion over
electron hopping around the atomic limit \cite{Metzner,ShvaikaPRB} and
allows to build DMFT in terms of local quantities. As a limit case the
Falicov-Kimball model with correlated hopping is considered.

In general, the hopping term of the Hamiltonian with correlated hopping
for the Falicov--Kimball model can be written as
\begin{equation}
  H_t=\frac{1}{\sqrt{D}}\sum_{<ij>}d_i^{\dag}d_j\Bigl[
  t_{ij}^{++}P_i^+P_j^+ + t_{ij}^{--}P_i^-P_j^-
  +t_{ij}^{+-}P_i^+P_j^- +
  t_{ij}^{-+}P_i^-P_j^+\Bigr],
  \label{HtFKG}
\end{equation}
where the projection operators $P_i^+=n_{if}$, $P_i^-=1-n_{if}$ on the
states of $f$-particles are introduced. It is convenient to rewrite it in
matrix notations
\begin{equation}\label{mtrx}
  H_t=\frac{1}{\sqrt{D}}\sum_{<ij>}{\mathbf d}_i^{\dag}{\mathbf t}_{ij}{\mathbf
  d}_j,\qquad
  {\mathbf d}_i=\left(\begin{array}{c} P_i^+ \\ P_i^- \end{array}\right) d_i, \qquad
  {\mathbf t}_{ij}=\left\|\begin{array}{cc}
                    t_{ij}^{++} & t_{ij}^{+-} \\
                    t_{ij}^{-+} & t_{ij}^{--}
                  \end{array}\right\|.
\end{equation}

The total Hamiltonian of the electronic system with correlated hopping
$H=\sum_i H_i + H_t$ includes besides the hopping term $H_t$ that is not
local also the single-site contributions $H_i$. Our aim is to consider the
$D\to\infty$ limit and it is convenient to start not from the Dyson
equation, that considers the terms with correlated hopping as some kind of
many-particle interactions, but from the Larkin equation
\cite{Larkin,Vaks} in coordinate representation
\begin{equation}\label{Larkin}
  {\mathbf G}_{ij}(\omega)={\mathbf \Xi}_{ij}(\omega)+
  \sum_{lm}{\mathbf \Xi}_{il}(\omega){\mathbf t}_{lm}{\mathbf
  G}_{mj}(\omega), \qquad
  {\mathbf G}_{{\mathbf k}}(\omega)=
  \left[{\mathbf \Xi}_{{\mathbf k}}^{-1}(\omega) - {\mathbf t}_{{\mathbf k}}\right]^{-1},
\end{equation}
that treats all hopping terms in a same manner. Here, an irreducible part
${\mathbf \Xi}_{ij}(\omega)$, that can not be divided into parts by
cutting one hopping line ${\mathbf t}_{lm}$, is introduced.

For the models with correlated hopping all quantities in (\ref{Larkin})
are matrices and the components of the Green's function ${\mathbf
G}_{ij}(\omega)$ are constructed by the projected (Hubbard) operators
\begin{equation}\label{GFcomp}
  G_{ij}^{\alpha\gamma}(\omega)=-\left\langle T P_i^{\alpha} d_i
  d_j^{\dag}P_j^{\gamma}\right\rangle_{\omega}=
  \beta\frac{\delta\Omega}{\delta t_{ji}^{\gamma\alpha}(\omega)},\qquad
  {\mathbf G}_{ij}(\omega)=-\left\langle T {\mathbf d}_i\otimes
  {\mathbf d}_j^{\dag}\right\rangle_{\omega},
\end{equation}
where $\Omega$ is the grand canonical potential functional, and the total
Green's function is equal
\begin{equation}\label{GFtot}
  G_{ij}(\omega)=-\left\langle T d_i d_j^{\dag}\right\rangle_{\omega}
  =\sum_{\alpha\gamma} G_{ij}^{\alpha\gamma}(\omega).
\end{equation}

In general, the irreducible part ${\mathbf \Xi}_{ij}(\omega)$ is
represented diagrammatically by the single-site vertices (because all
interactions in $H_i$ are local) connected by hopping
lines \cite{ShvaikaPRB} and in the $D\to\infty$ limit it can be shown (see,
e.g. Ref.~\cite{Metzner}) that all irreducible parts become local
\begin{equation}\label{Xiloc}
  {\mathbf \Xi}_{ij}(\omega)=\delta_{ij}{\mathbf \Xi}(\omega), \quad
  {\mathbf \Xi}_{{\mathbf k}}(\omega)={\mathbf \Xi}(\omega).
\end{equation}

Such matrix representation allows to reformulate the Dynamical Mean-Field
Theory of the systems with correlated hopping in terms of local
quantities. Indeed, the local irreducible part ${\mathbf \Xi}(\omega)$
depends on the electron hopping only through the local coherent potential
${\mathbf J}(\omega)=\sum_{lm}{\mathbf t}_{ol}\, {\mathbf
G}_{lm}^{[o]}(\omega)\, {\mathbf t}_{mo}$, where ${\mathbf
G}_{lm}^{[o]}(\omega)$ is the Green's function for the lattice with the
removed site $o$, and it is easy to show that coherent potential ${\mathbf
J}(\omega)$ is solution of the following equation
\begin{equation}\label{DMFTeq1}
  \frac1N\sum_{{\mathbf k}}
  \left[{\mathbf \Xi}^{-1}(\omega) - {\mathbf t}_{{\mathbf k}}\right]^{-1}
  =\left[{\mathbf \Xi}^{-1}(\omega) - {\mathbf J}(\omega)\right]^{-1}
  ={\mathbf G}_{{\mathrm imp}}(\omega)
\end{equation}
that is the matrix generalization of the known equation (see, e.g.
Ref.~\cite{DMFTreview}). Here, ${\mathbf G}_{{\mathrm imp}}(\omega)$ is
the Green's function for the effective single-impurity problem with the
statistical operator
\begin{equation}\label{rhoimp}
  \hat{\rho}_{{\mathrm imp}}={\mathrm e}^{-\beta H_o}T\exp\left\{\!
  -\!\int_0^{\beta}\!\!d\tau\!\int_0^{\beta}\!\!d\tau'
  {\mathbf d}_o^{\dag}(\tau){\mathbf J}(\tau-\tau'){\mathbf d}_o(\tau')
  \!\right\}.
\end{equation}

Finally, the grand canonical potential for the lattice can be expressed in
terms of the grand canonical potential for the impurity model
$\Omega_{{\mathrm imp}}$ by
\begin{equation}\label{latvsimp}
  \frac{\Omega_{{\mathrm lat}}}{N}=
  \Omega_{{\mathrm imp}}-\frac1{\beta}\sum_{\nu}
  \biggl\{\frac1N\sum_{{\mathbf k}}
  \ln\det\left[1-{\mathbf \Xi}(i\omega_{\nu})
  {\mathbf t}_{{\mathbf k}}\right]
  -\ln\det\left[1-{\mathbf \Xi}(i\omega_{\nu})
  {\mathbf J}(i\omega_{\nu})\right]
  \biggr\} .
\end{equation}

Now, let us apply the developed above approach to the Falicov--Kimball
model with correlated hopping. The single site Hamiltonian $H_i$ is
written
\begin{equation}\label{H_FK1}
  H_i=-\mu_f n_{if}-\mu_d n_{id}+Un_{id}n_{if}
  =-\mu_f P_i^+ + (U-\mu_d)n_{id}P_i^+ -\mu_d n_{id}P_i^-.
\end{equation}
Projection operators $P_i^+$ and $P_i^-$ commute with the total
Hamiltonian
%$H=\sum_i H_i + H_t$
and the partition function for the
impurity is a sum of the partition functions for the subspaces
$\alpha=\pm$:
\begin{equation}\label{Zimp}
  Z_{{\mathrm imp}}=\Sp\hat{\rho}_{{\mathrm imp}}={\mathrm e}^{-\beta Q_+} + {\mathrm e}^{-\beta
  Q_-},\qquad
  \Omega_{{\mathrm imp}}=
  -\frac1{\beta}\ln Z_{{\mathrm imp}},
\end{equation}
\begin{equation}
  Q_{\alpha} = -\mu_f\delta_{\alpha+} -
  \frac1{\beta}\ln\left(1+{\mathrm e}^{-\beta(U\delta_{\alpha+}-\mu_d)}\right)
         - \frac1{\beta}\sum_{\nu}\ln\left(1-
          \frac{J^{\alpha\alpha}(i\omega_{\nu})}
          {i\omega_{\nu}+\mu_d-U\delta_{\alpha+}}\right).
  \label{Qdef}
\end{equation}

The Green's functions matrix for the impurity is diagonal ($G_{{\mathrm
imp}}^{+-}(\omega)  = G_{{\mathrm imp}}^{-+}(\omega) = 0$)
\begin{equation}
  G_{{\mathrm imp}}^{++}(\omega)  =
  \frac{\langle P^+\rangle}{\omega+\mu_d - U - J^{++}(\omega)},\qquad
  G_{{\mathrm imp}}^{--}(\omega)  =
  \frac{\langle P^-\rangle}{\omega+\mu_d - J^{--}(\omega)},
  \label{GFimp}
\end{equation}
and, in addition, we have for the concentrations of the $f$ and $d$
particles
\begin{equation}\label{ndavg}
  n_f=\langle P^+\rangle=
  \frac1{Z_{{\mathrm imp}}}{\mathrm e}^{-\beta Q_+}, \qquad
  n_d=\langle d^{\dag}d\rangle=\frac1{\beta}\sum_{\nu}
  \left[G_{{\mathrm imp}}^{++}(i\omega_{\nu})
  +G_{{\mathrm imp}}^{--}(i\omega_{\nu})\right].
\end{equation}

Finally for the Green's function for the lattice we get
\begin{eqnarray}
  G_{{\mathbf k}}^{++}(\omega) &=&  \frac1{{\mathcal D}_{{\mathbf k}}(\omega)}
  \left[\omega + \mu_d - J^{--}(\omega)\langle P^+\rangle
  -t_{{\mathbf k}}^{--}\langle P^-\rangle\right]\langle P^+\rangle,
  \nonumber\\
  G_{{\mathbf k}}^{--}(\omega) &=&  \frac1{{\mathcal D}_{{\mathbf k}}(\omega)}
  \left[\omega + \mu_d - U - J^{++}(\omega)\langle P^-\rangle
  -t_{{\mathbf k}}^{++}\langle P^+\rangle\right]\langle P^-\rangle,
  \nonumber\\
  G_{{\mathbf k}}^{+-(-+)}(\omega) &=&  -\frac1{{\mathcal D}_{{\mathbf k}}(\omega)}
  \left[J^{+-(-+)}(\omega) - t_{{\mathbf k}}^{+-(-+)}\right]\langle P^+\rangle\langle P^-\rangle,
  \label{Glatfin}\\
  {\mathcal D}_{{\mathbf k}}(\omega)&=&
  \left[\omega+\mu_d-J^{--}(\omega)\langle P^+\rangle
  -t_{{\mathbf k}}^{--}\langle P^-\rangle\right]
  \nonumber\\
  &&\times\left[\omega+\mu_d-U-J^{++}(\omega)\langle P^-\rangle
  -t_{{\mathbf k}}^{++}\langle P^+\rangle\right]
  \nonumber\\
  &&-\left[J^{+-}(\omega) -t_{{\mathbf k}}^{+-}\right]
  \left[J^{-+}(\omega) -t_{{\mathbf k}}^{-+}\right]\langle P^+\rangle\langle
  P^-\rangle.
  \nonumber
\end{eqnarray}

Now, the total Green's function (\ref{GFtot}) for the Falicov--Kimball
model with correlated hopping can be written in the Dyson representation
as $G_{{\mathbf k}}(\omega)=\left[\omega+\mu_d-\Sigma_{{\mathbf
k}}(\omega)-\bar{t}_{{\mathbf k}}\right]^{-1}$, where $\bar{t}_{{\mathbf
k}}=t_{{\mathbf k}}^{++}\langle P^+\rangle^2 +t_{{\mathbf k}}^{--}\langle
P^-\rangle^2+ \left(t_{{\mathbf k}}^{+-}+t_{{\mathbf
k}}^{-+}\right)\langle P^+\rangle\langle P^-\rangle$ is the Hartree
renormalized (mean-field) hopping already introduced by Schiller
\cite{Schiller} and
\begin{equation}\label{Sigmatot}
  \Sigma_{{\mathbf k}}(\omega) = U\langle P^+\rangle
  +J_3(\omega) \langle P^+\rangle\langle P^-\rangle\\
   +\frac{U_{1{\mathbf k}}(\omega)U_{2{\mathbf k}}(\omega) \langle P^+\rangle\langle P^-\rangle}
  {\omega+\mu_d - U\langle P^-\rangle-\bar{J}(\omega)
  -t_{3{\mathbf k}}\langle P^+\rangle\langle P^-\rangle}
\end{equation}
is the momentum dependent (non local) self-energy. Here
\begin{eqnarray}
  U_{1(2){\mathbf k}}(\omega)&=& U
  +\left[t_{{\mathbf k}}^{++}-t_{{\mathbf k}}^{-+(+-)}-J^{--}(\omega)+J^{-+(+-)}(\omega)\right]
   \langle P^+\rangle
   \nonumber\\
  &&+\left[t_{{\mathbf k}}^{+-(-+)}-t_{{\mathbf k}}^{--}-J^{+-(-+)}(\omega)+J^{++}(\omega)\right]
   \langle P^-\rangle
   \label{Ueff}
\end{eqnarray}
describes the renormalization of the interaction $U$ and
\begin{eqnarray}\label{tJ}
  \bar{J}(\omega)&=&J^{++}(\omega)\langle P^-\rangle^2
  +J^{--}(\omega)\langle P^+\rangle^2
  +\left(J^{+-}(\omega)+J^{-+}(\omega)\right)\langle P^+\rangle\langle
  P^-\rangle,
  \nonumber\\
  J_3(\omega)&=&J^{++}(\omega)+J^{--}(\omega)-J^{+-}(\omega)-J^{-+}(\omega),
  \\
  t_{3{\mathbf k}}&=&t_{{\mathbf k}}^{++}+t_{{\mathbf k}}^{--}-
  t_{{\mathbf k}}^{+-}-t_{{\mathbf k}}^{-+}.
  \nonumber
\end{eqnarray}

In his article Schiller \cite{Schiller} considered the case of
$t_{3{\mathbf k}}=0$ and from (\ref{Sigmatot}) we get the same result for
the self-energy $\Sigma_{{\mathbf k}}(\omega)=\Sigma_0(\omega)+
\Sigma_1(\omega)t_{{\mathbf k}}+\Sigma_2(\omega)t_{{\mathbf k}}^2$ that
for the nearest-neighbor hopping $t_{{\mathbf k}}$ it contains only local,
nearest-neighbor and next-nearest-neighbor contributions. But in the
general case of $t_{3{\mathbf k}}\ne0$ the self-energy ``spreads'' over
all lattice.

Now, let us consider some limiting cases. At half filling $\mu_d=\mu_f$,
$n_f+n_d=1$, the spectral weight function is symmetric for two points
\cite{Schiller}. One of them correspond to the absence of correlated
hopping ($t^{\alpha\gamma}=t$). Another symmetric point $t_2/t_1=-1$,
$t_3=0$ ($t^{--}=t_1$, $t^{+-}=t^{-+}=t_1+t_2$, $t^{++}=t_1+2t_2+t_3$) was
considered for the 1D Hubbard model when exact results can be obtained
\cite{Strack}. In Ref.~\cite{Bulka} it was shown that in this case the
direct transition from the superconducting state to the Mott insulator
takes place. This symmetric point corresponds to the case of the diagonal
hopping matrix $t_{{\mathbf k}}^{--}=-t_{{\mathbf k}}^{++}=t_{{\mathbf
k}}$, $t_{{\mathbf k}}^{+-}=t_{{\mathbf k}}^{-+}=0$ when hopping is
allowed only between the sites with the same occupancy. Now, for the
Falicov--Kimball model the coherent potential matrix is also diagonal
($J^{+-}(\omega)=J^{-+}(\omega)=0$) and for the semi-elliptic density of
states of the half-width $W$ we have ($\alpha=\pm$)
\begin{equation}
  J^{\alpha\alpha}(\omega) = \frac{W^2}{4}G^{\alpha\alpha}(\omega)
  \label{Jpp}
  =\frac12\left(\omega+\mu_d - U\delta_{\alpha+}\right) +
  \frac i2\sqrt{W^2_{\alpha}-\left(\omega+\mu_d -
  U\delta_{\alpha+}\right)^2},
\end{equation}
where $W_{\pm}^2=W^2\langle P^{\pm}\rangle$. The spectral weight function
contains two bands
\begin{equation}\label{swf}
  \rho(\omega)=\frac1{\pi}\Img G_{{\mathrm imp}}(\omega-i0^+)
  = \frac{2\langle P^+\rangle}{\pi W_+^2}\sqrt{W_+^2-(\omega-U)^2}
  + \frac{2\langle P^-\rangle}{\pi W_-^2}\sqrt{W_-^2-\omega^2},
\end{equation}
separated by the gap $\Delta=U-W_+-W_-=U-W\left(\sqrt{\langle
P^+\rangle}+\sqrt{\langle P^-\rangle}\right)$ that is temperature
dependent out of half filling and at half-filling, when $n_f=n_d=\frac12$,
it disappears at $U_c=W\sqrt{2}$ (the Mott transition point).

\begin{figure}[h]
\begin{tindent}
\epsfxsize=0.45\textwidth\epsffile{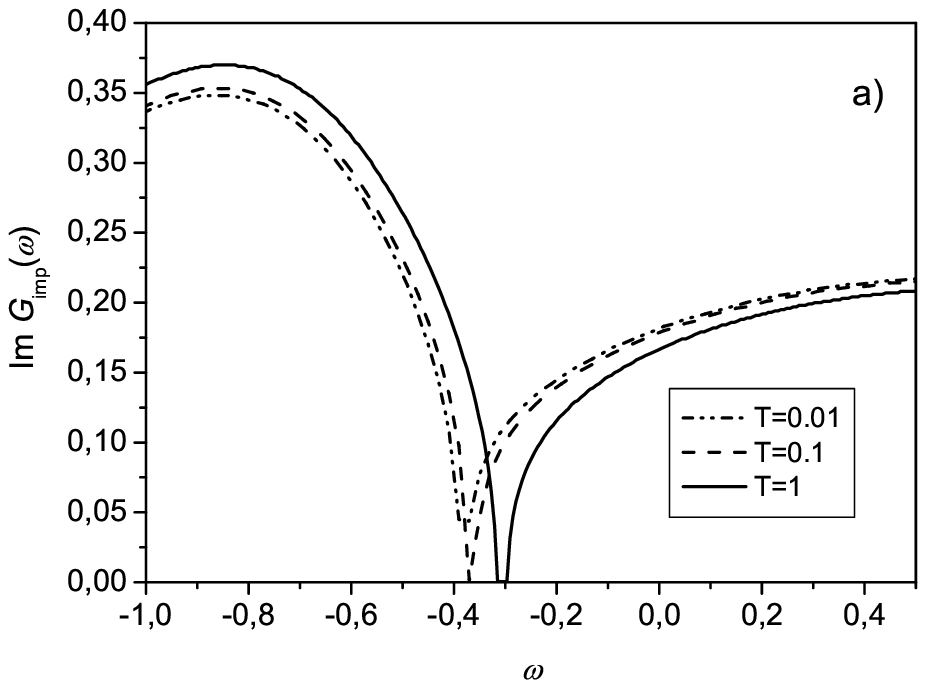}\quad
\epsfxsize=0.45\textwidth\epsffile{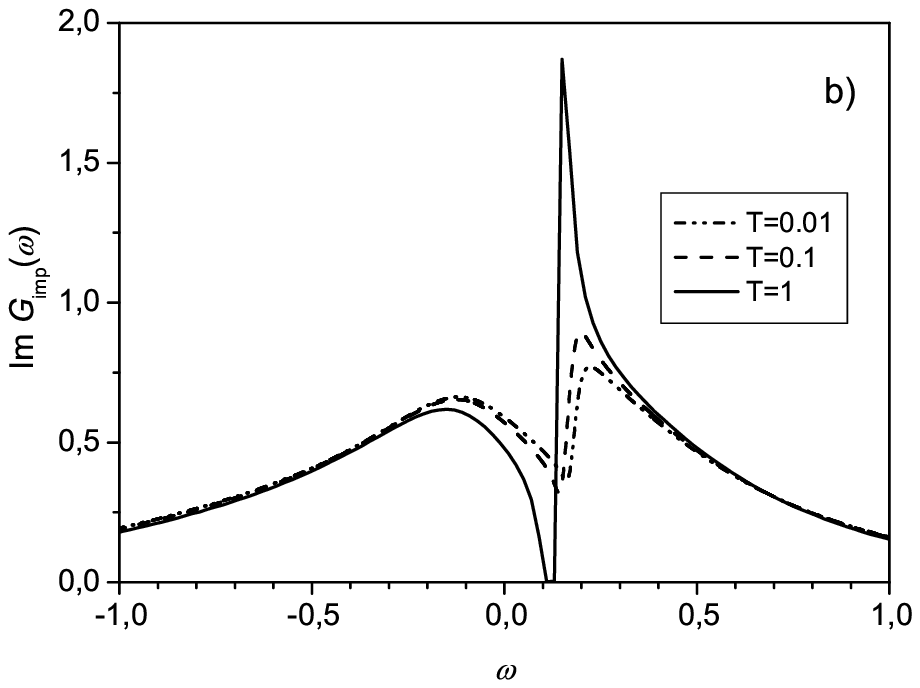}
\end{tindent}
\caption{Temperature development of the gap (Mott transition) for
$t_2/t_1=0.5$, $U=1.7$ (a) and $t_2/t_1=-0.5$, $U=0.3$ (b) for the
$D=\infty$ hypercubic lattice with nearest neighbor hopping at half
filling $\mu_d=\mu_f$, $n_f+n_d=1$}\label{fig3}
\end{figure}

Out of these symmetric points the Falicov--Kimball model with correlated
hopping possess the temperature driven Mott transition (see
Fig.~\ref{fig3}).

In this article we presented the general approach to the description of
the correlated hopping in the Dynamical Mean-Field Theory. It is based on
the Larkin equation (expansion over electron hopping around the atomic
limit) that considers all hopping terms, including correlated hopping, in
a same manner. Another starting point is the local character of the
irreducible part (irreducible cumulant) of the Green's functions
constructed by the projected (Hubbard) operators in the $D\to\infty$ limit
that is more general statement then the local character of the self-energy
which in the case of correlated hopping is unlocal. Such approach keeps
the Dynamical Mean-Field Theory local ideology and allows to calculate the
thermodynamical functions. As an example the Falicov-Kimball model with
correlated hopping and its limit case are considered when exact results
can be obtained. In particular, the temperature driven Mott transition in
the Falicov--Kimball model with correlated hopping is investigated.

This work was supported by the Science and Technology Center in Ukraine
under grant No~1673.

\end{document}